\documentclass[prl,reprint,showpacs]{revtex4-1}

\usepackage{graphicx}
\usepackage{amsfonts}
\usepackage{amssymb}
\usepackage{amsmath}
\usepackage{epstopdf}

\newcommand{\be}{\begin{equation}}
\newcommand{\ee}{\end{equation}}
\newcommand{\bea}{\begin{eqnarray}}
\newcommand{\eea}{\end{eqnarray}}

\begin{document}

\title{Trapping ultracold dysprosium: a highly magnetic gas for dipolar physics}

\author{Mingwu Lu}
\author{Seo Ho Youn}
\author{Benjamin L. Lev}
\affiliation{Department of Physics, University of Illinois at Urbana-Champaign, Urbana, IL 61801-3080 USA}

\begin{abstract}
Ultracold dysprosium gases, with a magnetic moment ten times that of alkali atoms and equal only to terbium as the most magnetic atom, are expected to exhibit a multitude of fascinating collisional dynamics and quantum dipolar phases, including quantum liquid crystal physics.  We report the first laser cooling and trapping of half a billion Dy atoms using a repumper-free magneto-optical trap (MOT) and continuously loaded magnetic confinement, and we characterize the trap recycling dynamics for bosonic and fermionic isotopes.  The first inelastic collision measurements in the few partial wave, 100 $\mu$K--1 mK, regime are made in a system possessing a submerged open electronic f-shell.  In addition, we observe unusual stripes of intra-MOT $<$ 10 $\mu$K sub-Doppler cooled atoms.  
\end{abstract}
\date{\today}
\pacs{37.10.De, 37.10.Gh, 37.10.Vz, 71.10.Ay}
\maketitle 

Ultracold gases of extraordinarily magnetic atoms, such as dysprosium, offer opportunities to explore strongly correlated matter in the presence of the long-range, anisotropic dipole-dipole interaction (DDI).  Such interactions in the presence (or absence) of polarizing fields can compete with short-range interactions to induce phases beyond those described by the nearest neighbor Hubbard model~\cite{Bloch:2008}.  Specifically, quantum liquid crystal (QLC) physics (see Ref.~\cite{Fradkin:2009} and citations within) describes strongly correlated systems in which a Fermi surface can spontaneously distort (nematics) or cleave into stripes (smectics)~\cite{Miyakawa:2008, *Fregoso:2009, *Quintanilla:2009}.  While material complexity can inhibit full exploration of QLC phases in condensed matter, QLC phases may be more extensively characterized in tunable ultracold gases.  In contrast to ultracold ground state polar molecules~\cite{Ye:2009}, ultracold Dy offers the ability to explore the spontaneously broken symmetries inherent in QLCs since the DDI is realized without a polarizing field.  An exciting prospect lies in observing spontaneous magnetization in dipolar systems, e.g., the existence of a quantum ferro-nematic phase in ultracold fermionic Dy gases not subjected to a polarizing field~\cite{Fregoso:2009b}. 

We report the first magneto-optical trap (MOT) and ultracold collisional rates of this highly complex atom.  The stable atoms possessing the largest magnetic moments are the neighboring lanthanide rare-earths, Tb and Dy (both 10 Bohr magnetons ($\mu_B$) to within 0.6\%~\cite{Martin:1978,Tb}).  Prior to the present work, the coldest Dy temperatures were achieved via buffer gas and adiabatic cooling to 50 mK with final densities of $<10^9$ cm$^{-3}$~\cite{Newman:2008}, and Dy beams have been transversely pushed and unidirectionally cooled via photon scattering~\cite{Leefer:2008,MetcalfBook99}.
\begin{figure}[t]
\includegraphics[width=0.49\textwidth]{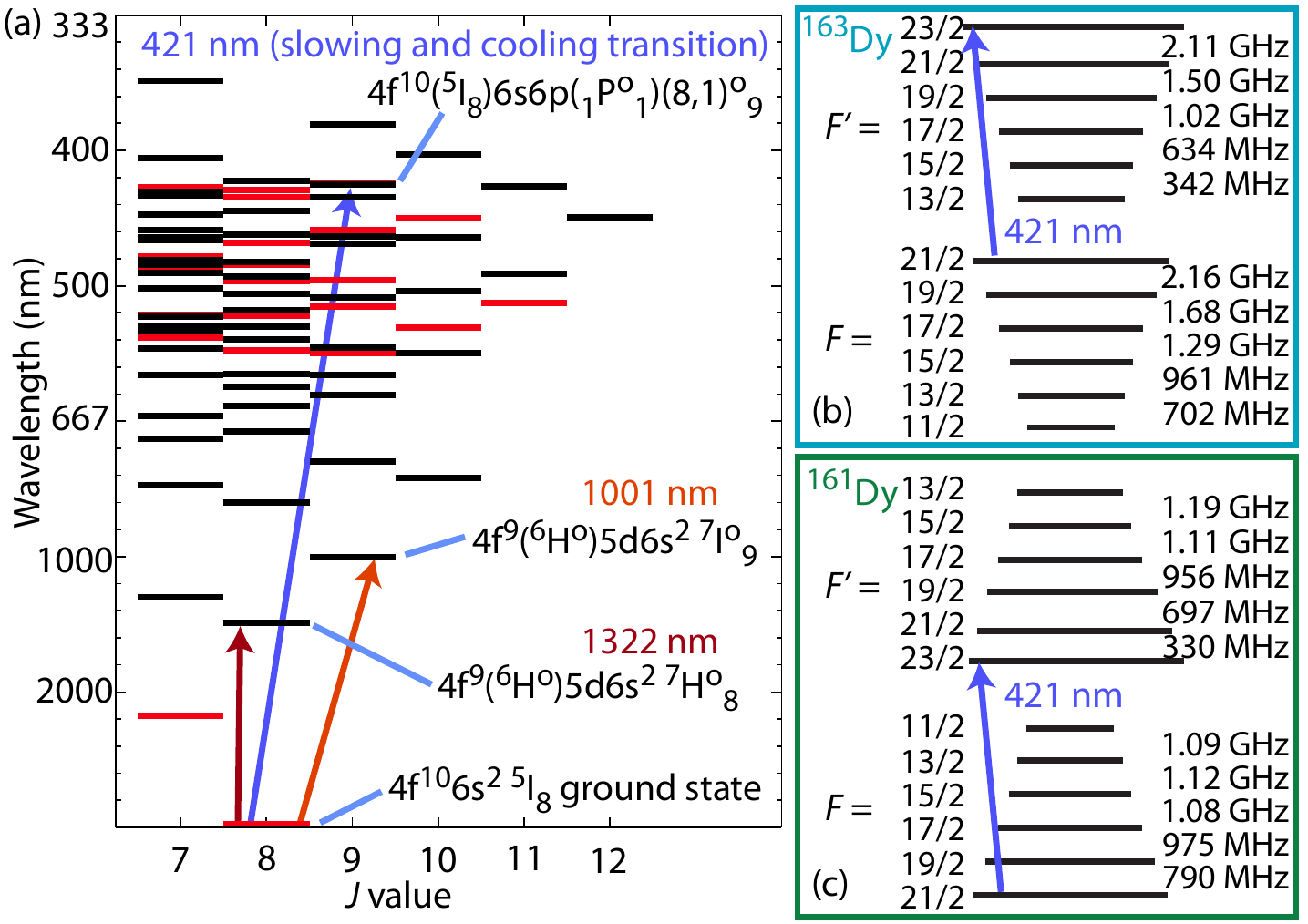}
\vspace{-7mm}
\caption{(color online). (a) Dysprosium energy level structure~\cite{Martin:1978,Leefer:2009}.  The MOT and Zeeman slower employ the strongest laser cooling transition between the even parity (red) ground state and the odd (black) excited state at 421 nm.  Dy has five high-abundance isotopes; three bosons ($^{164}$Dy, $^{162}$Dy, and $^{160}$Dy with $I=0$) and two fermions ($^{163}$Dy and $^{161}$Dy with $I=5/2$).  (b, c) Fermion hyperfine structure shown along with the cooling transition and energy splittings (not to scale).  Poor optical pumping reduces the $^{161}$Dy MOT population due to a large 14$\Gamma$ frequency difference between the $F$$=$$21/2$$\rightarrow$$F'$$=$$23/2$ and $F$$=$$19/2$$\rightarrow$$F'$$=$$21/2$ transitions.} 
\label{fig:dy_levels}
\vspace{-5mm}
\end{figure}

Recent experiments using degenerate $^{52}$Cr, a bosonic S-state atom with 6 $\mu_B$ of magnetic moment, have begun to explore quantum ferrofluids~\cite{Lahaye:2007}.  With suitable scattering lengths, the larger magnetic moment---and $9\times$ larger DDI$\cdot$mass ratio---of Dy should allow experimental access beyond the superfluid and Mott insulator regions of the extended Bose-Hubbard phase diagram to the density wave and supersolid regimes~\cite{Yi:2007}.  Co-trapping isotopes of Dy will allow exploration of dipolar Bose-Fermi mixtures of near equal mass.  Studies of degenerate spinor gases~\cite{Stamper-Kurn:2006,*Wu:2006} with large spins, and simulations of dense nuclear matter~\footnote{G. Baym and T. Hatsuda, private communication (2009).} are further exciting avenues of research.  

In addition, ultracold samples of Dy will aid precision measurements of parity nonconservation and variation of fundamental constants~\cite{Leefer:2008}, single-ion implantation~\cite{Mcclelland:2006}, and quantum information science~\cite{Derevianko:2004,*Saffman:2008}.  The low-lying telecommunications band (1322 nm) and InAs quantum dot (QD) amenable (1001 nm) transitions (Fig.~\ref{fig:dy_levels}) will enable hybrid quantum circuits of atom-photonic or atom-QD systems.  Novel collisional physics~\cite{Doyle:2009} and complex molecular association phenomena are expected in clouds of these $L\neq0$ atoms possessing submerged open f-shells inside closed 5s and 6s electron shells.

Dy structure is similar to the recently magneto-optically trapped Er (7 $\mu_B$)~\cite{Mcclelland:2006}.  Despite the multitude of excited state population loss channels, the repumper-less Dy MOT forms in a similar fashion to Er's:  The large magnetic moment allows Dy to remain magnetically trapped in the MOT region while excited state population decays through the metastable states.  No repumping lasers are necessary since a sufficient fraction of the atoms recycle to the MOT to overcome loss.  

\begin{figure}[t]
\includegraphics[width=0.49\textwidth]{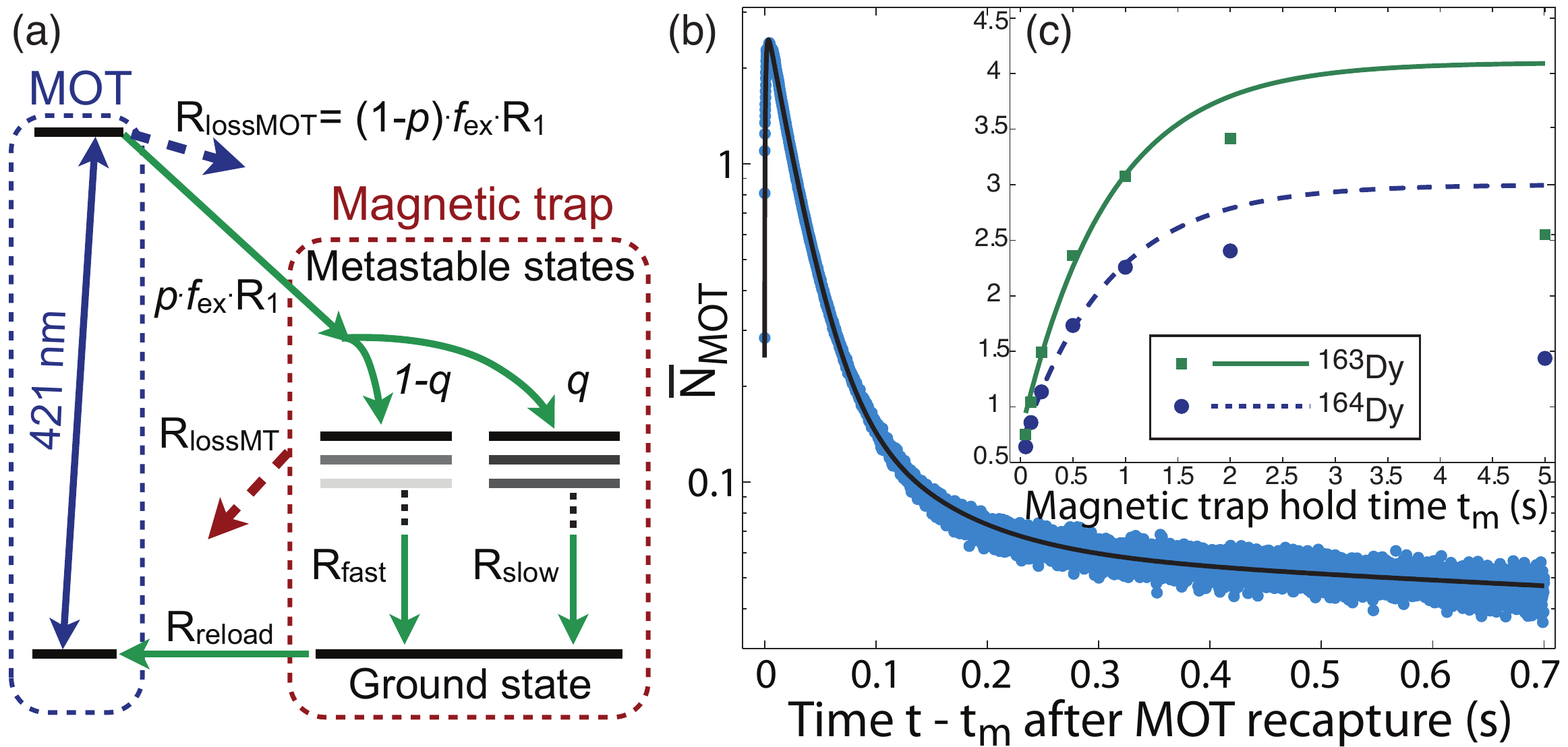}
\vspace{-6mm}
\caption{(color online). (a) Dy MOT recycling and continuously loaded magnetic trap (MT) schematic.  (b) Population ratio $\bar{N}_{\text{MOT}}$ of recaptured MOT to steady state MOT.  Black line is fit using Eqs.~\ref{MOTdecayeqns} to $\bar{N}_{\text{MOT}}$  with Zeeman slower and atomic beam off, and $t_m = 1$ s delay between steady state MOT and recapture.  
Inset (c) plots the peak recaptured MOT population versus hold time in MT. Lines are fits to $t\leq1$ s data with exponential time constants fixed at $R^{163}_{\text{slow}}=1.5$ s$^{-1}$ and $R^{164}_{\text{slow}}=2.3$ s$^{-1}$.}
\label{fig:MOTscheme}
\vspace{-5mm}
\end{figure}

The Dy cooling and trapping apparatus consists of a high temperature oven, transverse cooling stage, Zeeman slower, and MOT trapping region.  We elaborate on the design elsewhere~\cite{SeoHo:2009} with brief details here.  Nuggets of Dy are heated to 1250 $^{\circ}$C in a Ta crucible apertured to provide an atomic beam.   Three ion pumps---located at the oven, transverse pumping stage, and MOT chamber---along with titanium sublimation, achieve a vacuum of $1.8\times10^{-11}$ Torr during MOT operation.

The atomic beam passes through a differential pumping tube before intersecting four 100 mW, 2 cm-long transverse cooling beams detuned -0.2$\Gamma$ from the broad 421-nm transition~\footnote{Linewidth $\Gamma/2\pi=31.9\pm0.7$ MHz; see Refs.~\cite{SeoHo:2009,Martin:1978}}.  Transverse cooling enhances the optimized MOT population by a factor of 3--4.  The atomic beam then passes through a spin-flip Zeeman slower~\cite{MetcalfBook99} operating at -24$\Gamma$ detuning from the 421-nm line.  The power of the Zeeman slowing laser is $>$1 W at the slower entrance; we find a linear increase in MOT population until $\sim$1 W.  Atoms in the slowed beam are captured by a 6-beam MOT with detuning $\delta=-\Gamma$ from the 421-nm line and 12 mW in each 1.1-cm $e^{-2}$-waist beam (total intensity $I\,$$=\,$36 mW/cm$^2$).  We observe maximum MOT population for a 35 G/cm magnetic quadrupole gradient $\nabla_z B \approx 2\nabla_x B$.  The 421-nm lasers are derived from two doubled Ti:Sapphire lasers transfer-cavity locked to a Rb-stabilized diode laser.   

As the 421-nm laser system is detuned to the red of isotopes 160 through 164, steady state MOT populations form in proportion to natural abundance, except for the fermions $^{163}$Dy and $^{161}$Dy.  These isotopes' MOT populations are $5/6$$\times$ and $1/6$$\times$ the expected, respectively, which is likely due to poor optical pumping to the $F=21/2$ state~\cite{SeoHo:2009}, where $J$, $I$, and $F=J+I$ are the total electronic, nuclear, and total angular momenta.

Figure~\ref{fig:MOTscheme}(a) describes our model of the Dy MOT recycling mechanism, which is a refinement of that proposed for the Er MOT~\cite{Mcclelland:2006}.  The blue cooling laser excites a fraction $f_{\text{ex}}<1/2$ of the population to the 421-nm level, where it can decay with branching ratio $B = R_1/\Gamma$ to the metastable states.  Upon decay, a fraction $p f_{\text{ex}}$ of the population is captured in the magnetic quadrupole trap (MT) of the MOT at rate $p f_{\text{ex}}R_1$.  MOT recycling data, examples of which are shown in Figs.~\ref{fig:MOTscheme}(b) and (c), support a dual decay path through the dark metastable states; with probability $q$ ($1-q$) the population decays through the slow (fast) branch at rate $R_{\text{slow}}$ ($R_{\text{fast}}$).  Population that reaches the ground state reloads the MOT at rate $R_{\text{reload}}$, which depends on MOT parameters.  

There are two loss rates from the otherwise closed system (once loading from the Zeeman slower ceases).  A portion of the MOT population can decay to non-magnetically trapped metastable states at rate $R_{\text{lossMOT}}=(1-p)f_{\text{ex}}R_1$.  Additionally, metastable and ground state MT populations can be lost due to background and two-body inelastic collisions.  Two-body loss is difficult to quantify in the metastable states, but we investigate ground state MT loss below:  $R_{\text{slow}}$ is faster than the ground state component of $R_{\text{lossMT}}$, and we neglect $R_{\text{lossMT}}$ in the rate equations below.

\begin{figure}[t]
\includegraphics[width=0.49\textwidth]{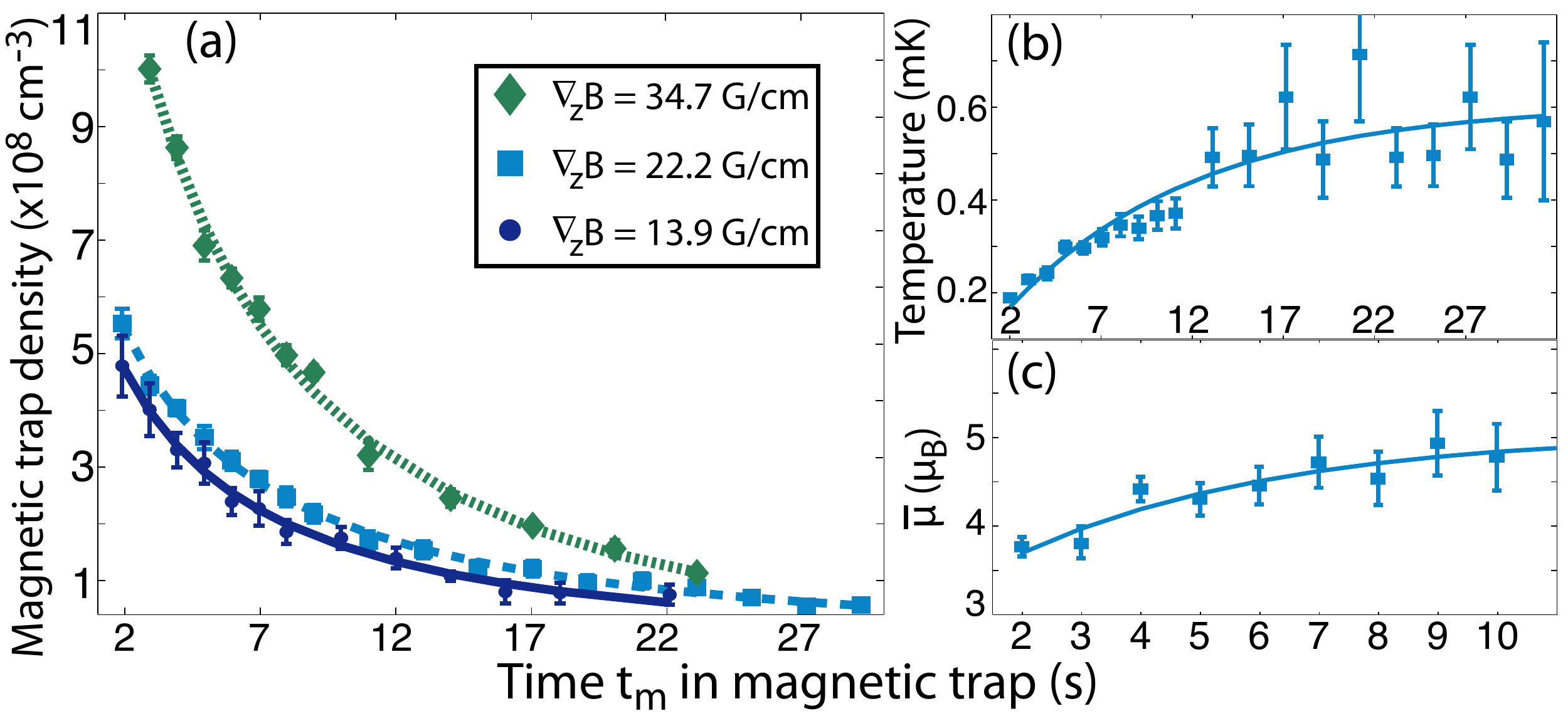}
\vspace{-7mm}
\caption{(color online). (a) Decay of magnetic trap mean density for three gradients.  Density profiles are consistent with a single thermal distribution after $t_m = 2$ s.  Data are fit to obtain one ($\gamma$) and two-body ($\beta$) collisional loss rates.  (b) Heating in $\nabla_zB=22.2$ G/cm magnetic trap.~(c) Increase in mean magnetic moment $\bar{\mu}$ for the same trap.  An exponential fit gives a $\bar{\mu}(t)$ time constant of $\tau=0.22(5)$ s$^{-1}$. }
\label{fig:MagTrap}
\vspace{-5mm}
\end{figure}

Following the procedure outlined in Ref.~\cite{Mcclelland:2006},  the rates in the MOT recycling model are determined by fitting the following equations to sets of MOT decay transients taken at various combinations of MOT beam intensity and detuning (see as an example the data in Fig.~\ref{fig:MagTrap}(b)):  
\bea\label{MOTdecayeqns}
\dot{N}_{\text{MOT}}&=&R_{\text{reload}}N_{\text{MT}}-f_{\text{ex}} R_1 N_{\text{MOT}}, \nonumber \\
\dot{N}_{\text{fast}}&=&(1-q)pf_{\text{ex}} R_1 N_{\text{MOT}}-R_{\text{fast}}N_{\text{fast}}, \nonumber \\
\dot{N}_{\text{slow}}&=&qpf_{\text{ex}} R_1 N_{\text{MOT}}-R_{\text{slow}}N_{\text{slow}}, \nonumber  \\
\dot{N}_{\text{MT}}&=&R_{\text{fast}}N_{\text{fast}}+R_{\text{slow}}N_{\text{slow}}-R_{\text{reload}}N_{\text{MT}},
\eea
where $N_{\text{MOT}}$, $N_{\text{fast}}$, $N_{\text{slow}}$, and $N_{\text{MT}}$ are the populations of the MOT, fast (slow) metastable state decay channel, and MT, respectively.  To obtain MOT recycling data with well defined initial conditions, the MOT is loaded with the Zeeman slowed atomic beam until reaching steady state, then the MOT, slower, and atomic beams are extinguished for $t_m=1$ s, during which most of the population in the metastable states decay to the ground state.  The population equations are numerically fit to the data for nine combinations of MOT power and detuning for the fermion $^{163}$Dy and eight for the boson $^{164}$Dy.  We verified that the hold time $t_m$ in the MT is sufficiently long that any residual population in $N_{\text{slow}}$ does not affect fit results; all other variables are left free to vary.  $R_1$ is extracted from the product $R = f_{\text{ex}} R_1$ by simultaneously fitting the $R$'s for each isotope to the function $R(I,\delta)=R_1\bar{I}/(2+2\bar{I}+2(2\delta/\Gamma)^2)$, where $\bar{I}=I/I_\text{s}$ and $I_\text{s}\approx2.7\times58$ mW/cm$^2$ in the MOT.  

Averaging the results, we find that for both isotopes
\bea
\left[R^{163}_1, R^{163}_{\text{fast}}, R^{163}_{\text{slow}}\right]
 &=&[1170(20),  19(2),1.5(1)]\text{ s}^{-1},  \nonumber\\
\left[R^{164}_1, R^{164}_{\text{fast}}, R^{164}_{\text{slow}}\right]
 &=& [1700(100),  29(1), 2.3(1)]\text{ s}^{-1}, \nonumber
\eea
and $[p,q]=[0.82(1), 0.73(1)]$.
The $R^{164}_1$ rate is consistent with the corresponding  bosonic $^{168}$Er ($I=0$) MOT quantity, and we suggest that the isotope-induced difference in Dy MOT decay rates arises from $^{163}$Dy's hyperfine structure.  Dy Zeeman slowing and MOT collection is possible because of the small $B\approx 7\times10^{-6}$~\footnote{$B$ is consistent with Er's and 4$\times$ smaller than that numerically estimated (private comm., V. Flambaum 2009)}.  The $p$ and $q$ values indicate that 82\% of the atoms are captured by the MT and 73\% of those cascade through the slow channel.  Dy level linewidths are mostly unknown, but we speculate that much of this population is captured by the low-lying and long-lived states near the 1322-nm telecom transition.

Using the measured $R_\text{slow}$ rates as fixed constants, data of the maximum MOT recapture population (peaks of data such as in Fig.~\ref{fig:MOTscheme}(b)) versus $t_m$ are fit to an exponential with only the initial and final MOT population as free variables.  Good fits are obtained for $t_m\leq1$ s data, as shown in Fig.~\ref{fig:MOTscheme}(c); we plot the fitted curve out to later times for comparison to data~\footnote{$\bar{N}_\text{MOT}<1$ at early times is due to the initial $N_\text{fast}\neq0$.}.  From these plots we can see that the population in the steady state MOT is only a fraction of the total number of trapped atoms.  The ``hidden" population in the continuously loaded MT is several times larger, filling the ground state MT at rate $R_\text{slow}$.  With respect to the steady state MOT, we measure nearly $2.5\times$ ($3.5\times$) more $^{164}$Dy ($^{163}$Dy) atoms in the recaptured MOT.  At optimal MOT parameters and after $t_m=2$ s of MT loading, we measure the total population of laser cooled and trapped atoms to be $5\times10^{8}$ for $^{164}$Dy and $3\times10^{8}$ for $^{163}$Dy.  Populations are measured via fluorescence collection on a fast photodiode with atom number calibration to $\leq$~10\% from absorption imaging~\cite{SeoHo:2009}.

Population loss at times $t_m>1$ s is likely due to background and inelastic collisions of ground state atoms in the MT. (Majorana spin flip losses are negligible at these $t_m$'s and trap densities.) We investigate these collision rates by absorption imaging the atoms confined in the MT at various $t_m$ delay times.  After loading the MOT, we extinguish the MOT, Zeeman slower, and atomic beams while maintaining a constant $\nabla B$.  The MT is turned off at $t_m$, and after 1 ms, an absorption image integrates on a 16-bit CCD camera for 200 $\mu$s.  Accounting for gravity $-g\hat{z}$, the images taken in the $x$--$z$ plane containing the quadrupole axis~(along $\hat{z}$) are fit to the functional form~\cite{Berglund:2007}:
\bea
p_x(x)&\propto&\exp(-2|x|\sqrt{1-\bar{g}^2}/\bar{x})(1+2|x|\sqrt{1-\bar{g}^2}/\bar{x}), \nonumber\\
p_z(z)&\propto&\exp(-2|z|/\bar{z}-2\bar{g}z/\bar{z})(1+2|z|/\bar{z}), \nonumber
\eea
where $\bar{g}=Mg/(\bar{\mu}\nabla_z B)$, $\bar{\mu}$ is the mean magnetic moment of the atomic cloud, and $\bar{x}=2k_BT/(\bar{\mu}\nabla_x B)$ (and similarly for $\bar{z}$).  We extract the mean cloud density $n$, temperature $T$, and $\bar{\mu}$ versus $t_m$, which we plot in Fig.~\ref{fig:MagTrap} for several MT gradients.  

The $^{164}$Dy density decay data in Fig.~\ref{fig:MagTrap}(a) are fit to the one plus two-body decay equation
$\dot{n}$=$-\gamma n$$-$$\beta n^2$,
where $\gamma$ is the loss rate due to background collisions and $\beta$ is a measure of inelastic two-body losses.  For the three trap gradients, the extracted $\beta$'s and $\gamma$'s are consistent with one another within 1$\sigma$; the weighted means are $[\bar{\beta},\bar{\gamma}]=[2.1(2)$$\times10^{-10}\text{ cm}{^3}/\text{s},5.6(3)$$\times10^{-2}\text{ s}^{-1}]$.  Fitting the data to $\dot{n}=- \beta n^2$ results in a worse $\chi^2$.  Temperature heating data---e.g., Fig.~\ref{fig:MagTrap}(b)---are fit to an exponential, resulting at early times to a heating rate of $17(2)$~$\mu$K/s, which is consistent with the heating rate in spin unpolarized Er and Cr MTs~\cite{Berglund:2007,Hensler:2003} and is likely due to spin-relaxation collisions.  The MT population is initially distributed among the weak-field seeking Zeeman states, and spin exchange collisions tend to polarize the sample toward $m_J = 8$.  Indeed, $\bar{\mu}$ increases with time, see Fig.~\ref{fig:MagTrap}(c), and $\bar{\mu}(t)$ reaches 8 $\mu_B$ within $t_m=4$~s in the high gradient $\nabla_zB=34.7$ G/cm trap. 

The measured $\bar{\beta}$ is likely due to an unresolved combination of inelastic spin exchange, magnetic dipole-dipole relaxation (MDDR), and anisotropic electrostatic-driven spin relaxation collisions.    Compared to $\beta_\text{Cr}=3$$\times10^{-11}$ cm$^{3}$/s in 200 $\mu$K spin-polarized $^{52}$Cr~\cite{Hensler:2003}, $\bar{\beta}$ is consistent with a MDDR scaling $\propto \mu^4$, though several times smaller than the full inelastic MDDR scaling presented in Ref.~\cite{Hensler:2003}.  In addition, $\bar{\beta}$ is consistent with non-maximally spin-polarized Cr (1.1$\times$10$^{-10}$ cm$^{3}$/s)~\cite{Hensler:2003} and with 500 mK inelastic Er--Er and Tm--Tm collision rates---3.0 and 1.1$\times10^{-10}$  cm$^{3}$/s, respectively~\cite{Doyle:2009}; the large magnitude of the latter rates are attributed to anisotropic electrostatic-driven spin relaxation collisions of these lanthanides.  Further measurements will aim to elucidate these collisional processes in this complex atom~\cite{SeoHo:2009}.

\begin{figure}[t*]
\includegraphics[width=0.49\textwidth]{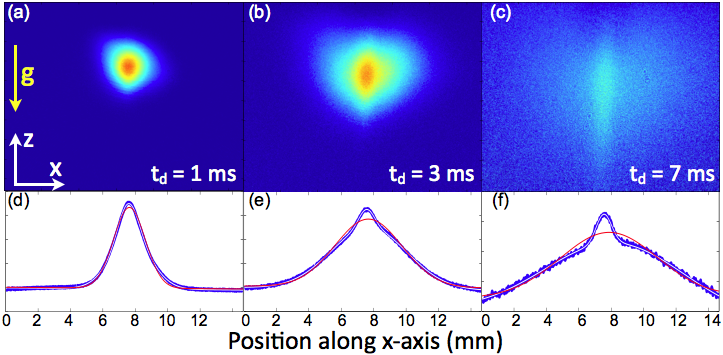}
\vspace{-6mm}
\caption{(color online). (a)-(c) $^{164}$Dy absorption images at time $t_d$ after release from MOT revealing a dense stripe with anisotropic temperature.  (d)-(f) Column density integrations along $\hat{z}$ versus $x$.  Fits to the data (blue) are to a double (white) and single (red) Gaussian. Additional images in the $x$-$y$ plane reveal the stripe's azimuthal symmetry.}
\label{fig:Temp}
\vspace{-5mm}
\end{figure}

Temperature and density profiles contain unusual features common to both $^{164}$Dy and $^{163}$Dy MOTs.  When care is taken to retroreflect and power balance all pairs of MOT beams, the MOTs are typically 30\% fewer in population, though have larger mean density (10$^{11}$ versus 10$^{10}$ cm$^{-3}$) than those with slight misadjustment.  In ballistic expansion after extinguishing the MOT, a dual component gas is observed comprised of a dense symmetric core of $\sim$200 $\mu$K atoms surrounded by a hot, 2--3 mK shell containing 70\% of the atoms. Doppler cooling theory in 1D predicts a cloud temperature of 1.2 mK for the MOT parameters, but comparisons to the Er MOT~\cite{Berglund:2007} suggest the entire intra-MOT population should be sub-Doppler cooled to $\sim$100 $\mu$K as a consequence of the near equal Land\'{e} $g$-factors in the ground and excited states ($\Delta g/g =$ 1.7\%).  Despite $\Delta g$ being slightly larger than in Er, numerical 1D sub-Doppler cooling simulations (based on Ref.~\cite{Berglund:2007} and references within) indicate that the entire Dy MOT---accounting for typical size and $\nabla B$---should be sub-Doppler cooled.  The Er MOT, limited by Zeeman and MOT laser power, contained 500$\times$~fewer atoms; it is possible that the hot shell forms subsequent to the cold inner core, which will be investigated.  At $t_m = 2$ s, MT temperatures are 100--300 $\mu$K depending on $\nabla B$.

While ultracold cores are observed in other MOTs~\cite{Jhe:2004}, a unique sub-Doppler cooled structure forms when the Dy MOT beams are slightly misbalanced or misaligned; an example of which is shown in Fig.~\ref{fig:Temp}.  Depending on the particular misadjustment, either a dense vertical or horizontal stripe (or both) appear in the core of the cloud.  The outer hot atoms remain at $T^H=2$--3 mK temperatures, but the (vertical) stripe population acquires an anisotropic temperature distribution of $T^C_z\approx T^H$ and $T^C_x\alt10$ $\mu$K.  This latter temperature is consistent the 1D numerical sub-Doppler cooling calculation, but its precise measurement is hampered by the hot cloud presence.  We offer no explanation other than to note no detection of stripe spin-polarization in Stern-Gerlach measurements (again hampered by the rapid hot cloud expansion), and that $T^C_x\rightarrow T^H$ as $\nabla B$ decreases~\cite{SeoHo:2009}.  Under optimal operating conditions, the Dy MOT phase space density is as large as $10^{-7}$.

Future work includes narrow-line cooling to $\sim$1 $\mu$K on the 1001-nm line~\cite{Berglund:2008} and loading into a crossed optical dipole trap.  Once optically confined, the large MT losses measured here might be avoided by trapping in the lowest energy Zeeman state.  Elastic and inelastic collision rates in the single partial wave regime will be measured before attempting to evaporatively cool to degeneracy.  The ultracold Dy produced with this method will open new avenues for research in quantum gases, precision measurement, and quantum information science.

\begin{acknowledgments}
We thank A{.} Berglund, U{.} Ray, J{.} McClelland, B{.} DeMarco, E{.} Fradkin, and J{.} Ye for technical assistance and critical reading.  We acknowledge support from the NSF, AFOSR, and ARO MURI on Quantum Circuits.
\end{acknowledgments}


\begin{thebibliography}{99}
\bibitem{Tb}%
  \BibitemOpen
  \bibinfo {note} {Unfortunately for exploring fermionic physics, Tb exists
  only as a single bosonic isotope. Tb also has a transition susceptible to
  incoherent 400 K blackbody radiation.}%
  \bibAnnoteFile{Stop}{Tb}%
\end{thebibliography}
%

\end{document}